\input{aipcheck}

\documentclass[
    ,final            
  ]
  {aipproc}

\layoutstyle{6x9}

\begin{document}

\title{Tau physics at the LHC with ATLAS}

\classification{14.60.Fg}
\keywords      {Tau lepton, ATLAS}

\author{Stan Lai\newline on behalf of the ATLAS Collaboration}{
  address={Albert-Ludwigs Universit\"{a}t Freiburg, 79104, Freiburg, Germany
           }
}

\begin{abstract}
The presence of $\tau$ leptons in the final state is an important signature in searches
for physics beyond the Standard Model.  Hadronically decaying
$\tau$ leptons can be reconstructed over a wide kinematic range at ATLAS.  
The reconstruction algorithm for hadronically decaying $\tau$
leptons and the performance of $\tau$ lepton identification is
described.  A review of physics processes with $\tau$ lepton final states is given, 
ranging from Standard Model processes in early data, such as $W$ and $Z$ boson production,
to searches for new phenomena beyond the Standard Model.
\end{abstract}

\maketitle


\subsection{Introduction}

The $\tau$ lepton, with a mass of $m_{\tau} = 1776.84\pm0.17$~MeV~\cite{PDG}, 
is the only lepton heavy enough to decay both leptonically and hadronically.  
It decays approximately 65\% of the time to one or more hadrons and 35\%
of the time leptonically.  The reconstruction and identification of $\tau$ leptons 
are important in many searches for new phenomena, and they can appear in final 
states in the production of Higgs bosons, supersymmetric (SUSY) particles, 
and other particles not described by the Standard Model~\cite{CSC, DetPap}.  
Standard Model processes, such as the $W$, $Z$ boson and $t\bar{t}$ production
can also result in signatures with $\tau$ leptons, and events from these
processes can be used to measure key quantities such as the 
$\tau$ lepton identification efficiency for the ATLAS 
reconstruction algorithm.

A challenge in identifying hadronically decaying $\tau$ leptons ($\tau_\mathrm{had}$) 
is to distinguish them from hadronic jets which are produced in processes 
with very large cross-sections.  However, $\tau_\mathrm{had}$ leptons possess 
certain properties that can be used to differentiate them from jets.  
They usually decay into one (1-prong) or three (3-prong) charged particles 
and their decay products are well collimated with an invariant mass less than 
$m_{\tau}$.  The $\tau$ lepton 
proper lifetime is 87~$\mu$m, leading to decay vertices that can be 
resolved in the silicon tracker from the primary interaction vertex.  
In addition, $\tau_\mathrm{had}$ leptons deposit a considerable fraction of 
their visible energy in the electromagnetic calorimeter in constrast with 
jets which deposit their energy primarily in the hadronic calorimeter.

\subsection{The reconstruction and identification of $\tau$ leptons}

Hadronically decaying $\tau$ candidates are reconstructed at ATLAS using 
two seed types.  The first seed type is a track with 
$p_\mathrm{T} > 6$~GeV that satisfies quality criteria on the number of 
associated silicon hits and the impact parameter with respect to the interaction vertex. 
The second type of seed consists of jets reconstructed using topological clusters 
(topoclusters)~\cite{CSC} with $E_T > 10$~GeV.  Topoclusters are formed using cells 
that exceed calorimeter noise by 4$\sigma$.  Neighbouring cells that exceed 
energy thresholds above calorimeter noise by 2$\sigma$ and 0$\sigma$ are associated 
to the cluster in a second and third step, respectively.  These topoclusters are then 
grouped into a topojet using a seeded cone algorithm~\cite{conealg} with a 
cone radius of $\Delta R = 0.4$ which 
forms seeds for $\tau_\mathrm{had}$ candidates.  
These topojets are then matched to the seed tracks in a cone of radius $\Delta R = 0.2$ 
around the topojet.   If such a match is found, the $\tau_\mathrm{had}$ candidate 
is considered as having two valid seeds.  For reconstructed $\tau_\mathrm{had}$ 
leptons in $Z\rightarrow\tau\tau$ events with $p_\mathrm{T} > 10$ GeV and $|\eta| < 2.5$, 
70\% of $\tau_\mathrm{had}$ candidates have two 
valid seeds, 25\% have only a topojet seed, and 5\% have only a track seed.

The energy of the $\tau_\mathrm{had}$ candidate is calculated in two ways.  
For $\tau_\mathrm{had}$ candidates with a topojet seed, the cells in a cone of 
$\Delta R = 0.4$ are summed and weighted according to a function dependent on 
the $\eta$, $\phi$ and calorimeter layer of the cell, similar to the method 
used for the Liquid Argon calorimeter of the H1 experiment~\cite{H1calib}.
For $\tau_\mathrm{had}$ candidates with a track seed, an energy flow approach is 
used, where energy deposits in cells matched to charged tracks are subtracted and 
replaced by the momenta of such tracks.  This energy of the $\tau_\mathrm{had}$ 
candidate is also corrected for energy leakage coming from charged particles 
outside the narrow cone.

Tracks associated to the $\tau_\mathrm{had}$ candidate in a cone of $\Delta R = 0.2$ 
are also required to pass track quality criteria on the number of associated hits 
in the silicon tracker and the impact parameter to the interaction vertex.  
Topoclusters found in the electromagnetic (EM) calorimeter with $E_\mathrm{T} > 1$ 
GeV that are isolated from tracks are interpreted as energy deposits from $\pi^0$ 
mesons in the $\tau$ lepton decay.  This procedure finds that approximately 66\% of 
$\tau \rightarrow \pi \nu$ decays are reconstructed with zero $\pi^0$ subclusters, 
while more than 50\% of $\tau \rightarrow \rho \nu$ ($\tau \rightarrow a_1 \nu$) 
decays are reconstructed with one (two) $\pi^0$ subcluster(s).

\begin{figure*}[h]
\centering
\includegraphics[width=64mm, height=50mm ]{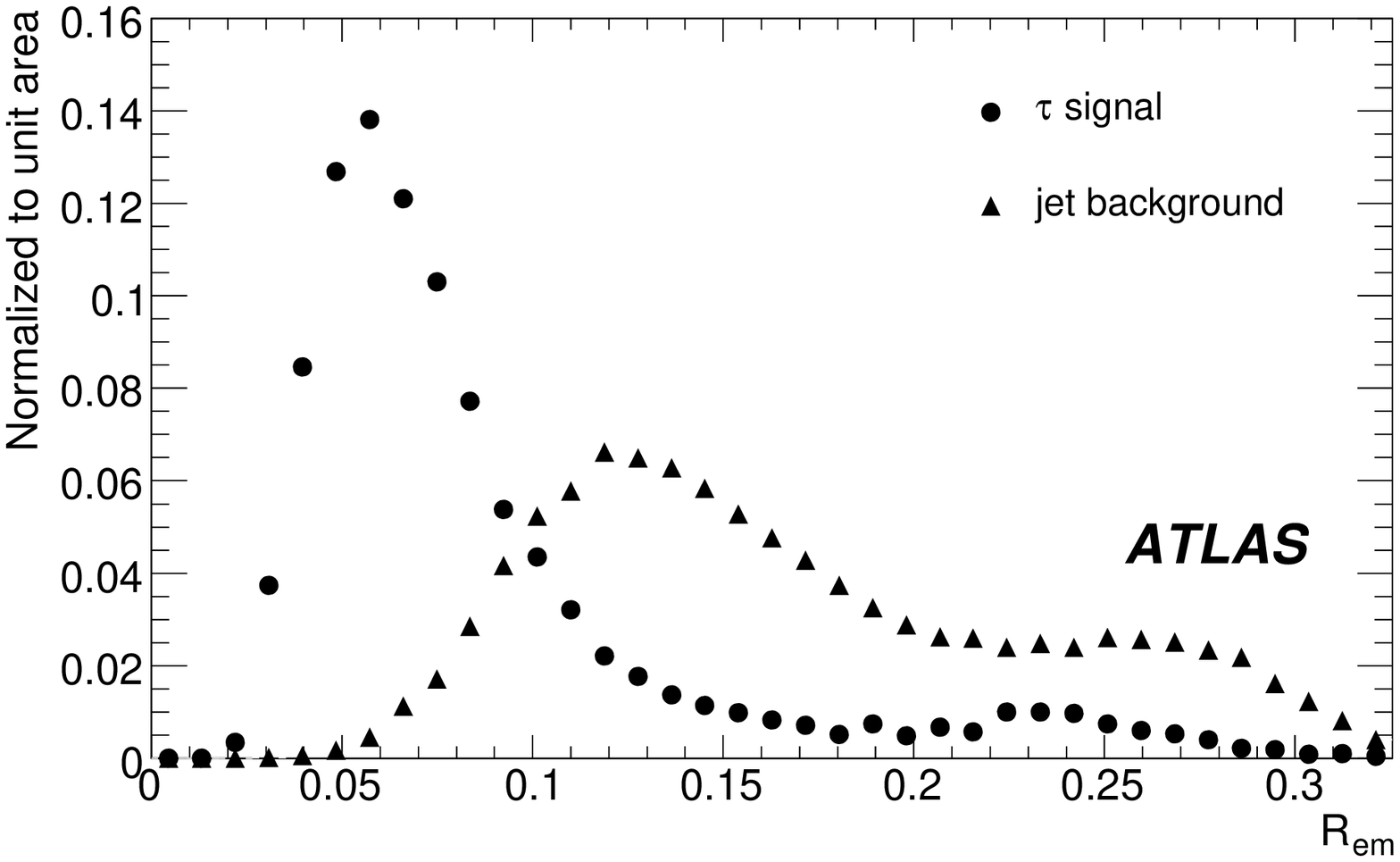}
~
\includegraphics[width=71mm]{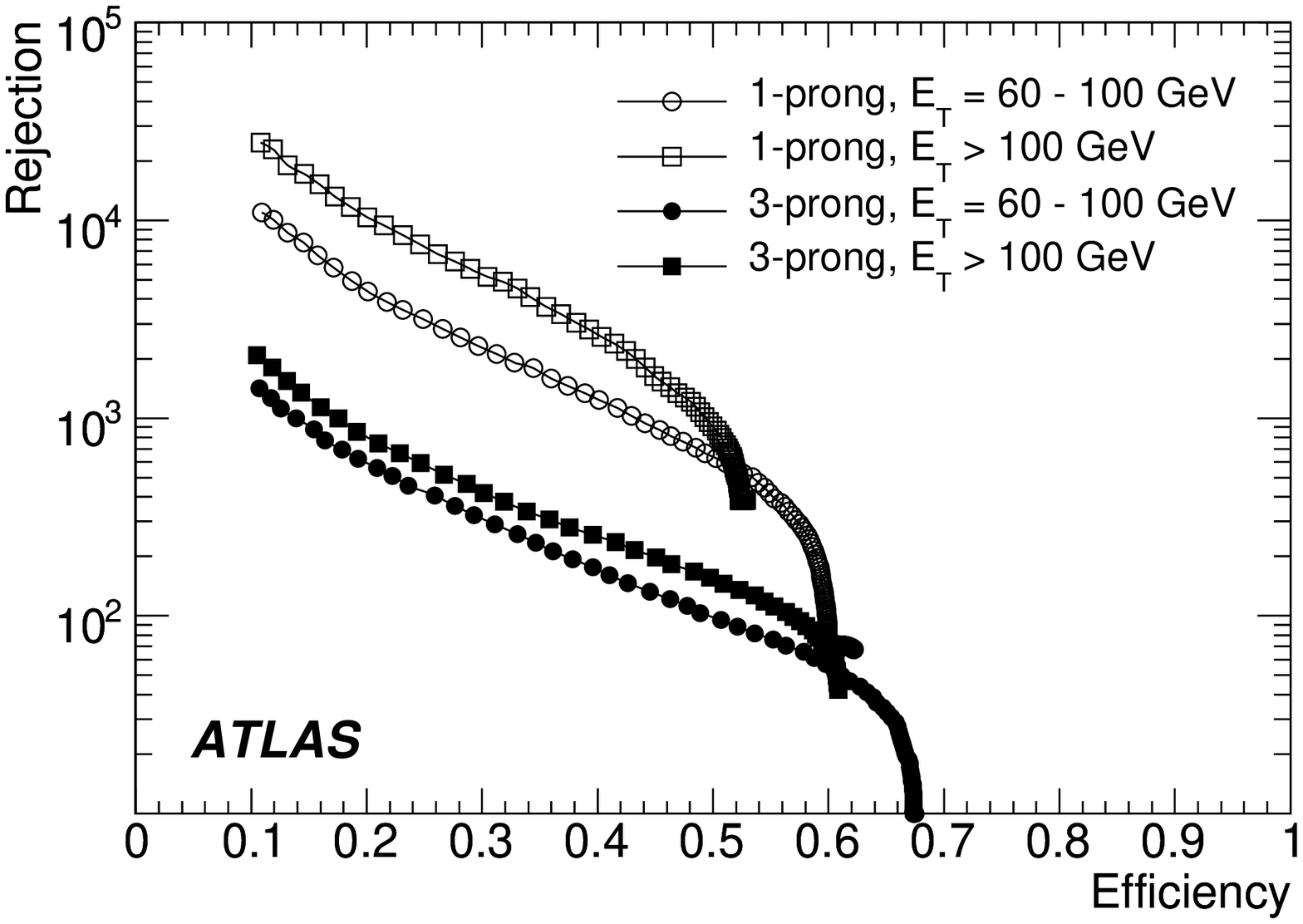}
\caption{ Left: The distribution of the reconstructed radius in the EM calorimeter ($R_\mathrm{EM}$)
                for $\tau_\mathrm{had}$ leptons and hadronic jets.
          Right: Rejection against jets as a function of $\tau$ lepton efficiency for the projective
                likelihood identification method for 1-prong and 3-prong candidates with
                $p_\mathrm{T}^{\tau} = 60-100$ and $>100$~GeV.
         } \label{alg}
\end{figure*}

Based on the calorimeter information, the associated tracks and 
reconstructed $\pi^0$ clusters, a variety of other variables are calculated to 
be used for the identification of $\tau_\mathrm{had}$ leptons.  These variables 
include the radius of energy deposits of the $\tau_\mathrm{had}$ candidate in the 
EM calorimeter (shown in Fig.~\ref{alg} [left]), isolation variables for the calorimeter 
energy and tracks, the reconstructed charge (based on the associated tracks), 
the invariant mass of the $\tau_\mathrm{had}$ candidate 
(with and without $\pi^0$ subclusters), the impact parameter significance of the 
leading track, ratios of energy deposits to the sum of track transverse momenta, 
and the transverse flight path significance of the $\tau_\mathrm{had}$ 
candidate vertex (for $\tau_\mathrm{had}$ candidates with more than one track).  

Muons are vetoed by requiring that the calorimetric energy deposited by the 
$\tau_\mathrm{had}$ candidate has $E_\mathrm{T} > 5$~GeV.  For electrons, cuts are 
placed upon $\tau_\mathrm{had}$ candidates on the following two quantities: the ratio of 
the transverse energy deposited in the EM calorimeter to the track transverse momentum 
which tends to be higher for electrons than for charged hadrons; and the ratio of high 
threshold hits to low threshold hits in the Transition Radiation Tracker for the track, 
which also tends to be higher for electrons.  This veto suppresses electrons 
by a factor of 60, while retaining 95\% of $\tau_\mathrm{had}$ leptons.

Many identification methods that use the reconstructed variables to suppress fake 
$\tau_\mathrm{had}$ lepton candidates from hadronic jets are studied at ATLAS, including cut-based 
identification, a projective likelihood, neural networks, and boosted decision trees.  
The rejection against jets as a function of identification efficiency for $\tau_\mathrm{had}$ leptons 
is shown in Fig.~\ref{alg} (right) for the projective likelihood for 1-prong and 3-prong candidates.

\subsection{Standard Model processes with $\tau$ final states}

The process $W \rightarrow \tau \nu$ has a cross-section of $1.7 \times 10^4$~pb at 
$E_\mathrm{CM} = 14$~TeV and it will be the most abundant source of $\tau$ leptons
at ATLAS.  Events from this process will be selected by a $\tau_\mathrm{had}$ lepton + 
$E_\mathrm{T}^\mathrm{miss}$ trigger, which has an efficiency of 70\% with respect to 
offline selection, and can only be run during luminosities less than 
$10^{32} \mathrm{cm}^{-2}\mathrm{s}^{-1}$.  Events are selected by requiring a 
$\tau_\mathrm{had}$ lepton with $20 < p_\mathrm{T} < 60$~GeV, an additional jet with 
$p_\mathrm{T} > 15$ GeV, and $E_\mathrm{T}^\mathrm{miss} > 60$~GeV not pointing 
in the direction of the jet or $\tau_\mathrm{had}$ lepton candidate.  Events that have 
an isolated electron or muon are also vetoed.

\begin{figure*}[h]
\centering
\includegraphics[width=90mm, height=60mm]{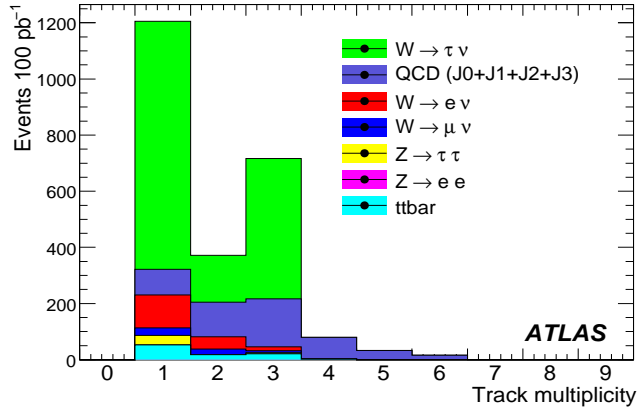}
\caption{The track multiplicity spectrum for $\tau_\mathrm{had}$ candidates in 
         100 pb$^{-1}$ for $W\rightarrow \tau \nu$ and background processes.
             } \label{Wplot}
\end{figure*}

The dominant background to this process is QCD dijet production, which has a production 
cross-section that is approximately six orders of magnitude greater than that for 
$W \rightarrow \tau \nu$.  Other backgrounds that contribute include $W \rightarrow e \nu$,
$W \rightarrow \mu \nu$, $t\bar{t}$, $Z \rightarrow \tau \tau$, and $Z \rightarrow ee$.
The signal to background ratio is very sensitive to the cut on $E_\mathrm{T}^\mathrm{miss}$.
A cut of $E_\mathrm{T}^\mathrm{miss} > 50$~GeV leads to a signal to background ratio of 1:1, 
while increasing this threshold to 60~GeV yields 3:1 for this ratio with 1550
signal events expected in 100~pb$^{-1}$ of integrated luminosity.  Therefore the scale and
systematic effects of $E_\mathrm{T}^\mathrm{miss}$ will have to be well under control.  
Fig.~\ref{Wplot} shows the track multiplicity spectrum for $\tau$ candidates in signal 
and background processes for 100~pb$^{-1}$ of integrated luminosity.

The $Z\rightarrow \tau\tau$ process will be the main channel of interest in early
data, despite having a cross-section approximately an order of magnitude less than
$W \rightarrow \tau \nu$.  The presence of an additional $\tau$ lepton can be 
used to further suppress backgrounds.  Studies involving this process usually
consider the case when one $\tau$ lepton decays hadronically 
and the other leptonically ($\tau_\mathrm{lep}$).  This also allows the events
to be selected using an electron or muon trigger, leading to an unbiased
sample of $\tau_\mathrm{had}$ candidates.

Backgrounds to $Z\rightarrow \tau\tau$ include dijet production, 
$W\rightarrow l\nu$ ($l = e,\mu,\tau$), $Z\rightarrow ee$, and $t\bar{t}$ events.
Events are selected by requiring an isolated electron or muon with 
$p_\mathrm{T} > 15$ GeV, an identified $\tau_\mathrm{had}$ lepton with
$p_\mathrm{T} > 15$ GeV, $E_\mathrm{T}^\mathrm{miss} > 20$~GeV, transverse mass
of the lepton and $E_\mathrm{T}^\mathrm{miss}$ satisfying 
$M_\mathrm{T}(\mathrm{lep}, E_\mathrm{T}^\mathrm{miss}) < 30$~GeV, 
sum of calorimeter energy deposits 
$\Sigma E_\mathrm{T}^\mathrm{calo} < 400$~GeV, and a veto on $b$-jets.
The lepton and $\tau_\mathrm{had}$ candidate must also be of
opposite sign, with requirements placed on 
$\Delta \phi(\mathrm{lep}, \tau_\mathrm{had})$ to ensure that they are not back-to-back.

\begin{figure*}[h]
\centering
\includegraphics[width=71mm]{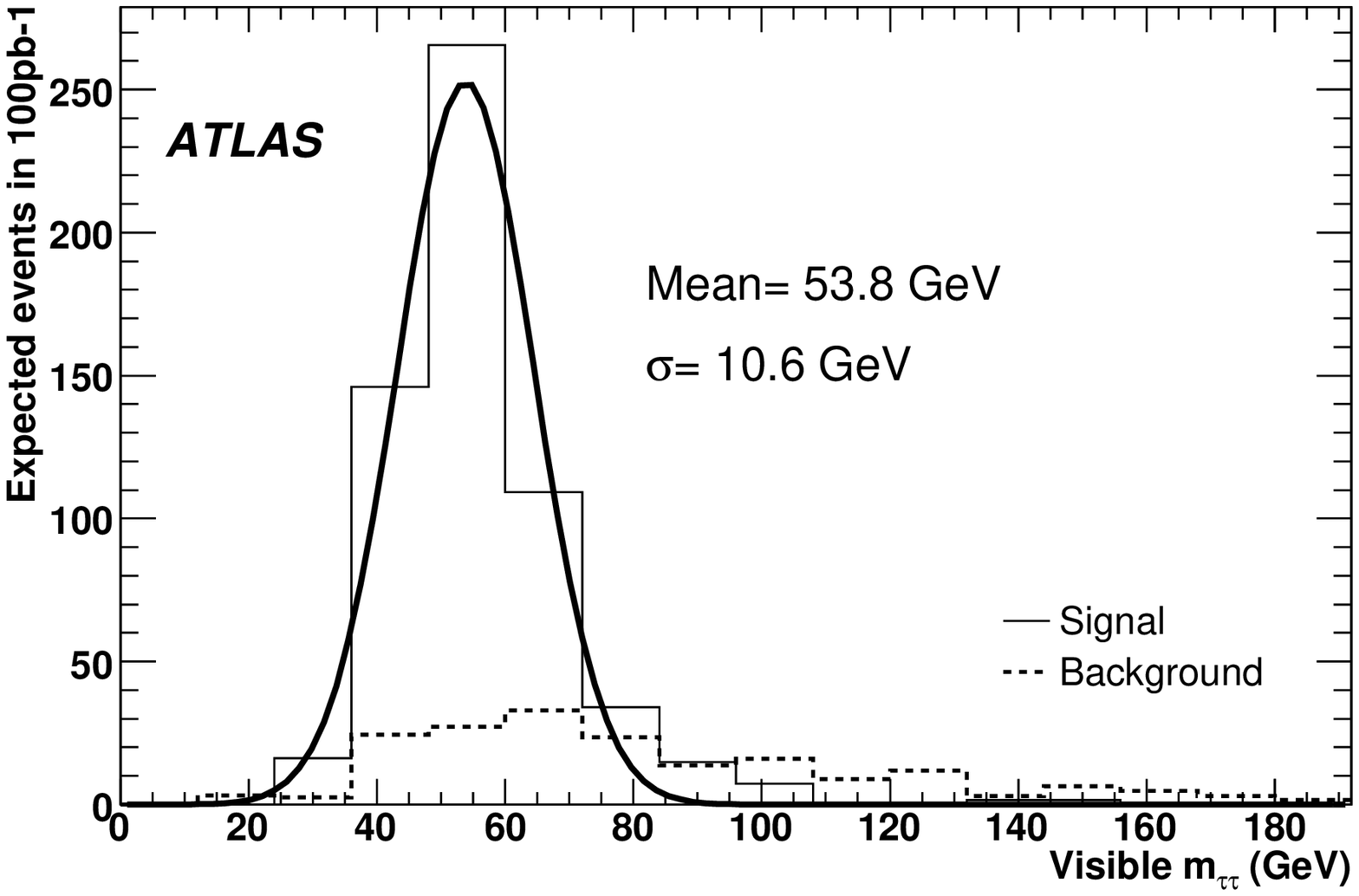}
~
\includegraphics[width=70mm]{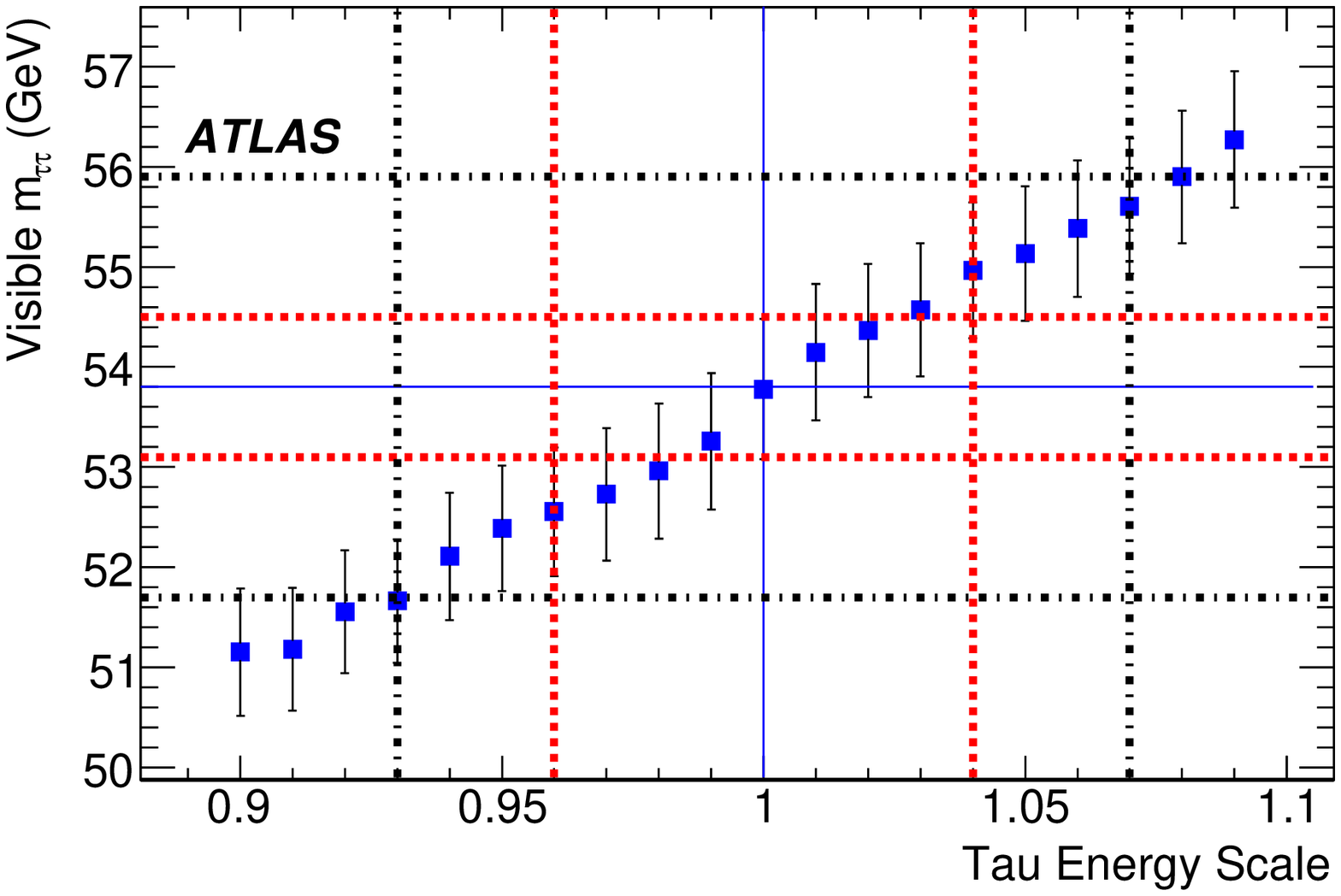}
\caption{Left: The visible mass $M_\mathrm{vis}(\mathrm{lep}, \tau_\mathrm{had})$ distribution 
               for $Z\rightarrow\tau\tau$ events and backgrounds in 100 pb$^{-1}$ of data.
         Right: The reconstructed visible mass $M_\mathrm{vis}(\mathrm{lep}, \tau_\mathrm{had})$
                as a function of the $\tau_\mathrm{had}$ energy scale (uncertainties shown
                are statistical only).   
        } \label{Zplot}
\end{figure*}

With 100~pb$^{-1}$, 520 events are expected with a visible mass 
$37 < M_\mathrm{vis}(\mathrm{lep}, \tau_\mathrm{had}) < 75$~GeV with a signal to background
ratio of 5:1.  The distribution of $M_\mathrm{vis}(\mathrm{lep}, \tau_\mathrm{had})$ is shown in
Fig.~\ref{Zplot} (left).  Backgrounds from dijet and $W$ production for this channel can be 
evaluated using data driven techniques, using control samples of same sign events, and the high  
$M_\mathrm{T}(\mathrm{lep}, E_\mathrm{T}^\mathrm{miss})$ region.

Since this channel provides an unbiased sample of $\tau_\mathrm{had}$ candidates,
tag and probe methods can be used to measure the identification efficiency for both
the offline reconstruction and trigger.  In addition, the visible mass distribution
can be used to evaluate the $\tau_\mathrm{had}$ lepton energy scale, shown
in Fig.~\ref{Zplot} (right).

\subsection{Searches for new phenomena with $\tau$ final states}

Many models that describe phenomena beyond the Standard Model predict signatures 
that are observable at the LHC with $\tau$ leptons in the final state.  For instance, some 
SUSY models predict an excess of events with $\tau$ leptons, particularly if the  
parameter $\tan \beta$ is large~\cite{SUSY}, when the $\tilde{\tau}$ is predicted to be the
next-to-lightest SUSY particle.

Strategies for SUSY searches with $\tau$ lepton final states have been evaluated assuming
1~fb$^{-1}$ of data, requiring a jet with $p_\mathrm{T} > 100$~GeV, three other jets with
$p_\mathrm{T} > 50$~GeV, a $\tau_\mathrm{had}$ candidate with $p_\mathrm{T} > 40$~GeV, 
$E_\mathrm{T}^\mathrm{miss} > 100$~GeV, and a veto on isolated electrons or muons.  The 
$E_\mathrm{T}^\mathrm{miss}$ is separated from the jets by requiring 
$\Delta R (E_\mathrm{T}^\mathrm{miss}, \mathrm{jet}) > 0.2$ and also must satisfy 
$E_\mathrm{T}^\mathrm{miss} > 0.2 \times M_\mathrm{eff}$ where the effective mass
$M_\mathrm{eff}$ is the scalar sum of the $p_\mathrm{T}$ of the jets, the 
$E_\mathrm{T}^\mathrm{miss}$ and $p_\mathrm{T}^{\tau_\mathrm{had}}$.  Additionally, 
the transverse sphericity greater than 0.2 is required, and a requirement of 
$M_\mathrm{T}(E_\mathrm{T}^\mathrm{miss}, \tau_\mathrm{had}) > 100$~GeV is also imposed. 

\begin{figure*}[h]
\centering
\includegraphics[width=90mm, height=62mm]{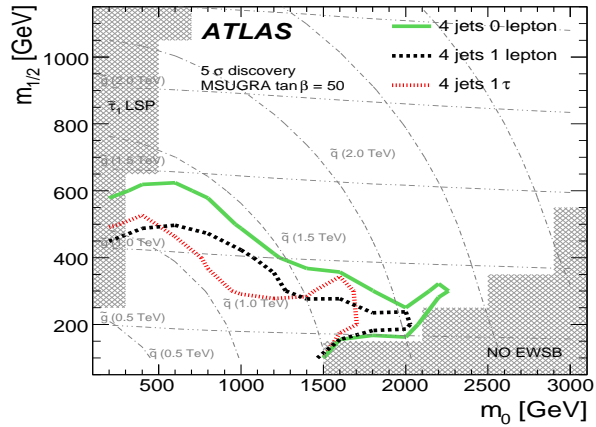}
\caption{ The $5\sigma$ discovery contours in the mSUGRA $m_0-m_{1/2}$ plane with $\tan \beta = 50$ 
          for 1~fb$^{-1}$.  The 4 jets + 1 $\tau_\mathrm{had}$ has similar sensitivity
          to the 4 jets search and the 4 jet + 1 lepton search.
             } \label{SUSYplot}
\end{figure*}

The main backgrounds for this channel include $t\bar{t}$, $W$ + jets, and $Z$ + jets
processes.  The discovery reach for this event selection in the mSUGRA $m_0-m_{1/2}$ plane 
with $\tan \beta = 50$ is shown in Fig.~\ref{SUSYplot} for 1~fb$^{-1}$ of integrated
luminosity.  The sensitivity of this search is comparable to similar search strategies 
without $\tau_\mathrm{had}$ lepton selection.

Other models posit the existence of heavier gauge bosons~\cite{BSM} such as the 
$Z^\prime$ boson, with some predicting enhanced couplings to $\tau$ leptons.  The most 
promising prospects occur in the $Z^\prime \rightarrow \tau_\mathrm{had}\tau_\mathrm{lep}$ 
channel.  In this case, an isolated electron (muon) with $p_\mathrm{T} > 27 (22)$~GeV, 
an identified $\tau_\mathrm{had}$ lepton with  $p_\mathrm{T} > 60$~GeV and opposite charge,
$E_\mathrm{T}^\mathrm{miss} > 30$~GeV and 
$M_\mathrm{T}(\mathrm{lep}, E_\mathrm{T}^\mathrm{miss}) < 35$~GeV are required. 
The vector sum of $E_\mathrm{T}^\mathrm{miss}$, 
$p_\mathrm{T}^{\tau_\mathrm{had}}$ and $p_\mathrm{T}^\mathrm{lep}$ must be less than
70~GeV, and a visible mass cut of $M_\mathrm{vis} (\mathrm{lep}, \tau_\mathrm{had}) > 300$ 
GeV is applied.  
To extract the invariant mass of the $\tau\tau$ system, the collinear approximation is
used, which infers the direction of neutrinos from $\tau$ decay by projecting the 
components of the $E_\mathrm{T}^\mathrm{miss}$ on the axes defined by the visible 
$\tau$ lepton decay products.
A cut on $\cos\Delta\phi (\mathrm{lep}, \tau_\mathrm{had}) > -0.99$ ensures 
that the lepton and $\tau_\mathrm{had}$ candidates are not back-to-back so that the
collinear approximation can be applied.

\begin{figure*}[h]
\centering
\includegraphics[width=77mm, height=57mm]{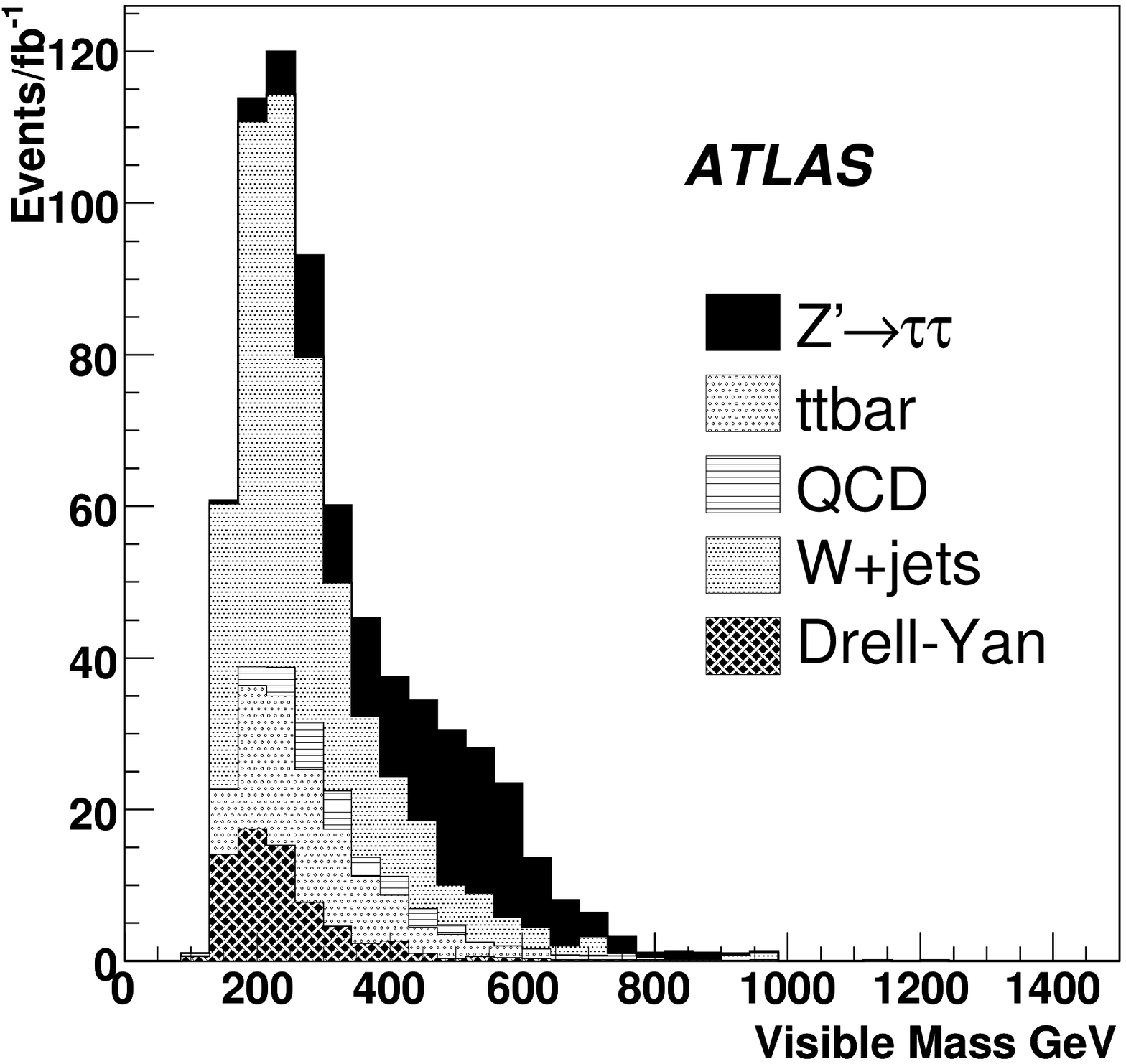}
~
\includegraphics[width=77mm, height=57mm]{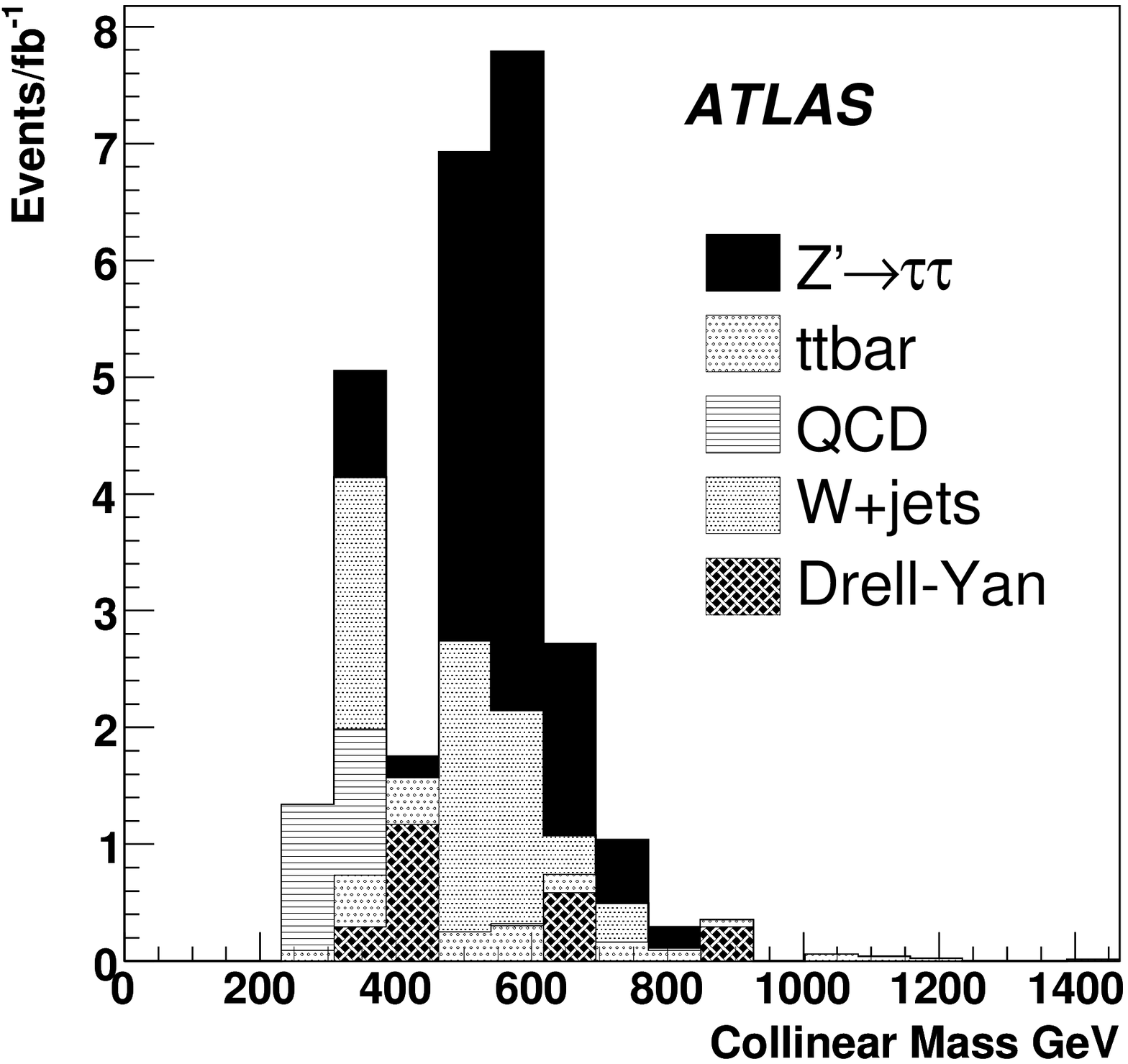}
\caption{ Left: The visible mass $M_\mathrm{vis} (\mathrm{lep}, \tau_\mathrm{had})$ distribution
                for $Z^\prime$ events and associated backgrounds ($m_{Z^\prime} = 600$~GeV).
          Right: The invariant mass $M_\mathrm{inv} (\mathrm{lep}, \tau_\mathrm{had})$ distribution 
                 attained from the collinear approximation for $Z^\prime$ events and 
                 associated backgrounds ($m_{Z^\prime} = 600$~GeV). 
             } \label{Zprimeplot}
\end{figure*}

The distributions for the visible mass $M_\mathrm{vis} (\mathrm{lep}, \tau_\mathrm{had})$
and invariant mass $M_\mathrm{inv} (\mathrm{lep}, \tau_\mathrm{had})$ attained through the collinear
approximation are shown in Fig.~\ref{Zprimeplot}, where Standard Model couplings
for the $Z$ boson are extended to the $Z^\prime$ boson, and the mass $m_{Z^\prime} = 600$~GeV
is assumed.  A clear excess of $Z^\prime$ events can be seen over Standard Model
backgrounds.  Systematic uncertainties are dominated by uncertainties in the 
integrated luminosity acquired and the $\tau_\mathrm{had}$ lepton energy scale.

\subsection{Conclusions}

The ATLAS experiment has developed a reconstruction algorithm for hadronically 
decaying $\tau$ leptons that exploits their well known properties to suppress fake
$\tau_\mathrm{had}$ candidates coming from hadronic jets, electrons or muons.  Standard
Model processes like $W$ and $Z$ boson production will be an abundant source of $\tau$ 
leptons, enabling early data measurements of $\tau_\mathrm{had}$ lepton identification, 
$\tau_\mathrm{had}$ energy scale, and cross-section measurements of $W$ and $Z$ boson 
production in $\tau$ lepton channels.  Once $\tau_\mathrm{had}$ lepton reconstruction 
is well understood in data, promising searches for new phenomena beyond the Standard 
Model with $\tau$ lepton final states can be performed.


\begin{thebibliography}{9}

\bibitem{PDG}
C. Amsler {\it et al.} (Particle Data Group), Physics Letters B667, 1 (2008).

\bibitem{CSC}
The ATLAS Collaboration,  G. Aad {\it et al.}, ``Expected Performance of the ATLAS Experiment, Detector, Trigger,
and Physics'', CERN-OPEN-2008-020, Geneva, (2008).

\bibitem{DetPap}
The ATLAS Collaboration, G. Aad {\it et al.}, ``The ATLAS Experiment at the CERN Large Hadron
Collider'', JINST 3 (2008) S08003.

\bibitem{conealg}
S.D. Ellis {\it et al.}, ``Jets in hadron-hadron collisions'', Prog.Part.Nucl.Phys. 60:484-551, (2008).

\bibitem{H1calib}
C. Schwanenberger (for the H1 Collaboration), ``The Jet Calibration in the H1 Liquid Argon Calorimeter'', 
arXiv:physics/0209026v1 [physics.ins-det] (2002).

\bibitem{SUSY}
S.P. Martin, ``A Supersymmetry Primer'', arXiv:hep-ph/9709356v5, (1997).

\bibitem{BSM}
H. Georgi, S. Weinberg, ``Neutral currents in expanded gauge theories'', Phys. Rev. D, \textbf{17}, 275, (1978).

\end{thebibliography}
\end{document}